\documentstyle[12pt,aasms4]{article}

\def\kms{km~s$^{-1}$}
\def\ref{\reference{}}

\received{}
\accepted{}
\journalid{}{}
\articleid{}{}
\lefthead{Deaton et~al.}
\righthead{Spectra of SN 1994I}

\begin{document}

\title{Direct Analysis of Spectra of the Type~Ic Supernova 1994I}

\author{{JENNIFER~MILLARD\altaffilmark{1}},
{DAVID~BRANCH\altaffilmark{1}}, {E.~BARON\altaffilmark{1}},
{KAZUHITO~HATANO\altaffilmark{1}}, {ADAM~FISHER\altaffilmark{1}},
{ALEXEI~V.~FILIPPENKO\altaffilmark{2}}, {R.~P.~KIRSHNER\altaffilmark{3}},
{P.~M.~CHALLIS\altaffilmark{3}}, {CLAES FRANSSON\altaffilmark{4}},
{NINO PANAGIA\altaffilmark{5}}, {M.~M.~PHILLIPS\altaffilmark{6}},
{GEORGE SONNEBORN\altaffilmark{7}}, {N.~B.~SUNTZEFF\altaffilmark{6}},
{R.~V.~WAGONER\altaffilmark{8}}, and {J.~C.~WHEELER\altaffilmark{9}}}

\altaffiltext{1}{Department of Physics and Astronomy, University of
Oklahoma, Norman, OK~73019}

\altaffiltext{2}{Department of Astronomy, University of California,
Berkeley, CA~94720--3411}

\altaffiltext{3}{Harvard--Smithsonian Center for Astrophysics,
60~Garden St., Cambridge, MA~02138}

\altaffiltext{4}{Stockholm Observatory, S--133~36 Saltsj\"obaden,
Sweden}

\altaffiltext{5}{Space Telescope Science Institute, 3700 San Martin
Drive, Baltimore, MD~21218 (on assignment from Space Science
Department of ESA)}

\altaffiltext{6}{CTIO, NOAO, Casilla~603, Le~Serena, Chile}

\altaffiltext{7}{Laboratory for Astronomy and Solar Physics,
NASA/GSFC, Code~681, Greenbelt, MD~20771}

\altaffiltext{8}{Department of Physics, Stanford University,
Stanford, CA~94305}

\altaffiltext{9}{Department of Astronomy, University of Texas,
Austin, TX~78712}

\begin{abstract}

Synthetic spectra generated with the parameterized supernova
synthetic--spectrum code SYNOW are compared to observed
photospheric--phase spectra of the Type~Ic supernova 1994I.  The
observed optical spectra can be well matched by synthetic spectra that
are based on the assumption of spherical symmetry.  We consider the
identification of the infrared absorption feature observed near
10,250~\AA, which previously has been attributed to He~I
$\lambda10830$ and regarded as strong evidence that SN~1994I ejected
some helium.  We have difficulty accounting for the infrared
absorption with He~I alone.  It could be a blend of He~I and C~I
lines.  Alternatively, we find that it can be fit by Si~I lines
without compromising the fit in the optical region.

In synthetic spectra that match the observed spectra, from 4 days
before to 26 days after the time of maximum brightness, the adopted
velocity at the photosphere decreases from 17,500 to 7000 \kms.
Simple estimates of the kinetic energy carried by the ejected mass
give values that are near the canonical supernova energy of $10^{51}$
ergs.  The velocities and kinetic energies that we find for SN~1994I
in this way are much lower than those that we find elsewhere for the
peculiar Type~Ic SNe~1997ef and 1998bw, which therefore appear to have
been hyper--energetic.

\end{abstract}

 
\section{Introduction}

Supernova~1994I, in the Whirlpool Galaxy (M51$\equiv$NGC~5194), was a
well observed supernova of Type~Ic.  By definition, the spectrum of a
Type~Ic lacks the strong hydrogen lines of a Type~II, the strong He~I
lines of a Type~Ib, and the deep red Si~II absorption of a Type~Ia
(e.g., Filippenko (1997).  A Type~Ic (SN~Ic) is thought to be the
result of the core collapse of a massive star that either loses its
helium layer before it explodes, or ejects some helium that remains
insufficiently excited to produce conspicuous optical He~I lines.  In
the specific case of SN~1994I, the identification of an absorption
feature observed near 10,250~\AA\ with the He~I $\lambda10830$ line
(Filippenko et~al. 1995) has been taken to indicate that helium was
ejected.

Recently, interest in SNe~Ic has been very high, because of the
peculiar SNe~Ic 1998bw and 1997ef.  SN~1998bw probably was associated
with a gamma--ray burst (Galama et~al. 1998; Kulkarni et~al. 1998;
Pian 1999; but see Norris, Bonnell, \& Watanabe 1998).  It also was
exceptionally bright at radio wavelengths and relativistic ejecta
seem to be required (Kulkarni et~al. 1998; Li \& Chevalier 1999).  On
the basis of optical light--curve studies, the kinetic energy of its
ejected matter has been inferred to be much higher than the canonical
supernova energy of 10$^{51}$ ergs (Iwamoto et~al. 1998a; Woosley,
Eastman, \& Schmidt 1998), unless the matter ejection was extremely
asymmetric (Wang \& Wheeler 1998).  Spectroscopically, SN~1997ef
resembled SN~1998bw, but Iwamoto et~al. (1998b) were able to fit the
SN~1997ef light curve with a normal kinetic energy.  However, on the
basis of synthetic--spectrum studies such as that which is presented
in this paper for SN~1994I, we (Deaton et~al. 1998; Branch 1999;
Millard et~al. in preparation) find that SN~1997ef, like SN~1998bw, was
hyper--energetic.  The light curve of SN~1997ef can also be fit with a
high kinetic energy (K.~Nomoto, private communication.)

According to light--curve studies of SN~1994I (Nomoto et~al. 1994;
Iwamoto et~al. 1994; Young, Baron, \& Branch 1995; Woosley, Langer, \&
Wheeler 1995), it was {\sl not} a hyper--energetic event .  Thus, in
addition to being of interest in its own right as the best observed
``ordinary'' SN~Ic, SN~1994I provides a valuable basis for comparison
with the apparently hyper--energetic SNe~1997ef and 1998bw.  It should
be noted that the spectra of SNe~1998bw and 1997ef were definitely
unusual; they differed significantly from those of normal SNe~Ic such
as SN~1994I, but a comparative study (Branch~1999) suggests that the
main difference is that SNe~1997ef and 1998bw ejected more mass at
high velocity.

In this paper we report the results of a study of photospheric--phase
spectra of SN~1994I using the parameterized supernova
spectrum--synthesis code SYNOW (Fisher et~al. 1997, 1999).  The
observed spectra are discussed in Section~2 and our
synthetic--spectrum procedure is described in Section~3. Our results
are presented in Section~4 and discussed in Section~5.  In a companion
paper (Baron et~al. 1999), detailed NLTE synthetic spectra of some
hydrodynamical models are compared with SN~1994I spectra.

\section{Observations}

Figure~1 displays the spectra that we have studied with SYNOW. The
spectra have been corrected for interstellar reddening using $A_V=1.2$
mag.  It is clear that the reddening was substantial, but the amount
is uncertain [e.g., Richmond et~al. (1996) estimated $A_V=1.4\pm0.5$
mag]. Throughout this paper, spectral epochs are in days with respect
to the date of maximum brightness in the $B$ band, 1994 April~8~UT
(Richmond et~al. 1996).  The six spectra of Figure~1~(a) were obtained
at the Lick Observatory by Filippenko et~al. (1995), who presented
additional spectra that are not reproduced here; see their paper for
details of the observations and reductions.  Also in Figure~1~(a), a
spectrum obtained by the {\sl Supernova INtensive Study} (SINS) group
with the {\sl Hubble Space Telescope} ({\sl HST}) at $+11$ days is
combined with the Filippenko et~al. (1995) optical spectrum for $+10$
days. The seven spectra of Figure~1~(b) were obtained by Brian Schmidt
and R.P.K. at the Multiple Mirror Telescope; only two of these been
published previously, in Baron et~al. (1996).  Photospheric--phase
optical spectra of SN~1994I also have been published by Sasaki
et~al. (1994; spectra obtained from $-5$ to $+8$ days) and Clocchiatti
et~al. (1996; from $-2$ to $+56$ days).

The most detailed discussions of line identifications in SN~1994I, and
SNe~Ic in general, have been by Clocchiatti et~al. (1996, 1997).  The
major contributors to some of the spectral features in Figure~1 are
clear.  The Ca~II H and K doublet ($\lambda\lambda3934,3968$) and the
Ca~II infrared triplet ($\lambda\lambda8498,8542,8662$) are
responsible for the P Cygni features at $3600-4000$~\AA\ and
$8000-8800$~\AA, respectively.  The Na~I D lines
($\lambda\lambda5890,5892$) produce the feature at $5500-6000$~\AA\
(and interstellar Na~I in M51 produces the narrow absorption near
5900~\AA).  O~I $\lambda7773$ produces the feature at $7400-
7900$~\AA.  The structure from about 4300 to 5400~\AA\ is recognizable
as blended Fe~II lines.

The line identifications of the other features --- including those in
the {\sl HST} ultraviolet and those longward of 8800~\AA --- are less
obvious.  As mentioned above, the strong absorption near 10,250~\AA,
which we will refer to as ``the infrared absorption'', has previously been
identified with He~I $\lambda10830$ but we will consider other
possible contributors.

Figure~1 shows at a glance that, as is usual for supernovae, the
absorption features tend to drift redward with time, as line formation
takes place in ever deeper, slower layers of the ejected matter.
Anticipating the results of the analysis presented in Section~3, we
show in Figure~2 the velocity at the photosphere, $v_{phot}$, that we
adopt for our spectral fits.  Our adopted values of $v_{phot}$ fall
from 17,500 \kms\ at $-4$ days to 7000 \kms\ at $+26$ days.

\section{Procedure}

We use the fast, parameterized, supernova spectrum--synthesis code
SYNOW to make a ``direct'' analysis (Fisher et~al. 1997, 1999) of
spectra of SN~1994I.  The goal is to establish line identifications
and intervals of ejection velocity within which the presence of lines
of various ions is detected, without adopting any particular
hydrodynamical model.  The composition and velocity constraints that
we obtain with SYNOW then can provide guidance to those who compute
hydrodynamical explosion models and to those who carry out
computationally intensive non--local--thermodynamic--equilibrium
(NLTE) spectrum modeling.  The SYNOW code is described briefly by
Fisher et~al. (1997) and in detail by Fisher (1999).  In our work on
SN~1994I we have made use of the paper by Hatano et~al. (1999), which
presents plots of LTE Sobolev line optical depths versus temperature
for six different compositions that might be expected to be
encountered in supernovae, and displays SYNOW optical spectra for 45
individual ions that can be regarded as candidates for producing
identifiable features in supernova spectra.  (Electronic data from the
Hatano et~al. [1999] paper, now extended from the optical to include
the infrared, can be obtained at the website
www.nhn.ou.edu/$\sim$baron/papers.html.)

For comparison with each observed spectrum in Figure~1 we have
calculated many synthetic spectra with various values of the following
fitting parameters.  The parameter $T_{bb}$ is the temperature of the
underlying blackbody continuum; the values that we use range from
18,000~K at $-4$ days to 6000~K at $+26$ days.  The parameter
$T_{exc}$ is the excitation temperature.  For each ion that is
introduced, the optical depth of a reference line is a fitting
parameter, and the optical depths of the other lines of the ion are
calculated for Boltzmann excitation at $T_{exc}$.  Little physical
significance is attached to $T_{exc}$; even if Boltzmann excitation
held strictly, the adopted value of $T_{exc}$ would represent some
kind of average over the line forming region.  The values that we use
range from 12,000~K to 5000~K.  The radial dependence of the line
optical depths is taken to be a power law, $\tau \propto v^{-8}$, and
the line source function is that of resonance scattering. The most
interesting fitting parameters (and therefore the only ones that we
quote for each individual synthetic spectrum) are velocity parameters.
The values that we use for $v_{phot}$, the velocity of matter at the
photosphere, have been plotted in Figure~2.  The outer edge of the
line forming region is at $v_{max}$, which we have fixed at 40,000
\kms. In addition, we can introduce restrictions on the velocity
interval within which each ion is present; when the minimum velocity
assigned to an ion is greater than $v_{phot}$, the line is said to be
detached from the photosphere.

\section{Results}

Although we have fit all of the spectra in Figure~1, only the fits to
the Lick spectra (and the single {\sl HST} spectrum) are shown here, because
they have more extended wavelength coverage than the MMT spectra.
Instead of discussing the spectra in chronological order, we first
discuss those obtained from $+7$ to $+26$ days, and then work backward
in time from $+4$ to $-4$ days.  One reason for doing this is that
establishing line identifications generally is more difficult at
earlier times, when line formation takes place in higher velocity
layers and the blending is more severe.

\subsection{From $+7$ to $+26$ days}

Figure~3 compares the $+7$ day observed spectrum to a synthetic
spectrum that has $v_{phot}=10,000$ \kms. The synthetic spectrum
includes lines of C~II, O~I, Na~I, Ca~II, Ti~II, and Fe~II, and the
fit is good.  (The excessive height of the synthetic peaks in the blue
part of the spectrum is not of great concern; the number of lines
having significant optical depth rises very rapidly toward short
wavelengths [e.g., Wagoner, Perez, \& Vasu 1991] so our SYNOW
synthetic spectra often are underblanketed in the blue because of
missing weak lines of unused ions.)  The identifications of Ca~II,
O~I, Fe~II, and Na~I are definite. Lines of Ti~II have been introduced
to fit the absorption near 4200~\AA, and we consider Ti~II to be
positively identified also.  Lines of C~II also are introduced, but
detached at 16,000 \kms, so that $\lambda6580$ can account for most of
the observed absorption near 6200~\AA.  At some epochs it is difficult
to decide between detached C~II $\lambda$6580 and the undetached Si~II
$\lambda\lambda$6347,6371 doublet (cf. Fisher et~al. 1997), but at
this epoch Si~II would produce an absorption too far to the blue of
the observed one.  We consider the C~II identification to be probably,
but not definitely, correct.  The main spectral features that remain
to be explained, then, are those around 7000 and 9000~\AA, and the IR
absorption.

The $+7$ day observed spectrum appears again in Figure~4, where it is
compared to three more synthetic spectra.  The one in the upper panel
is like the one shown in Figure~3, except that (1) He~I lines have
been introduced in an attempt to account for the core of the IR
absorption, and (2) the C~II lines have been removed to allow the
effects of the optical He~I lines to be seen more clearly.  For He~I,
instead of making our usual approximation that the relative
populations of the lower levels of interest are given by LTE at
excitation temperature $T_{exc}$, we assume that the populations of
the lower levels of the optical He~I lines (2$^3$P, 2$^1$P) are
further reduced, relative to the population of the lower level of
$\lambda10830$ (2$^3$S), by the geometrical dilution factor (which has
the value 0.5 at the photosphere and decreases with radius).  This is
a reasonable approximation that is roughly consistent with the results
of detailed calculations for He~I by Lucy (1991; his Figure~3) and
Mazzali \& Lucy (1998; their Figures~1 and 8).  Even with this
reduction in the optical depths of the optical He~I lines relative to
that of $\lambda10830$, when we try to account for the entire infrared
absorption with $\lambda10830$ the synthetic features produced by the
optical lines, $\lambda$7065 (2$^3$P), $\lambda$5876 (2$^3$P), and
$\lambda$6678 (2$^1$P), are much too strong.  As shown in the upper
panel of Figure~4, even when we attempt to account only for the {\sl
core} of the infrared absorption with $\lambda10830$, by reducing the
He~I optical depth and detaching the He~I lines at 15,000 \kms, the
optical He~I lines are still too strong. Thus we find that it is
difficult to account for the IR absorption with He~I $\lambda10830$
alone.

Lines of C~I have been discussed in connection with the infrared feature by
Woosley \& Eastman (1997) and Baron et~al. (1996).  We find that like
He~I $\lambda10830$, C~I (multiplet 1, $\lambda10695$) cannot
account for the entire infrared absorption without compromising the fit in
the optical.  But, as shown in the middle panel of Figure~4, we find
that undetached C~I $\lambda10695$, combined with He~I $\lambda10830$
detached at 18,000 \kms, can do so.  C~I also accounts for the
observed absorption near 9300~\AA\ and it helps near 6800~\AA,
although it does produce an absorption that is stronger than the
observed one near 8800~\AA.  Introducing C~I allows the optical depth
of He~I $\lambda10830$ to be reduced to the extent that the optical
He~I lines do no harm.

To complicate matters, however, the lower panel of Figure~4 shows that
lines of Si~I (mainly multiplets 4, $\lambda$12047; 5, $\lambda$10790;
6, $\lambda$10482; and 13, $\lambda$10869), detached at 14,000 \kms,
can account for the entire infrared absorption while also doing some good in
the optical, especially near 7000~\AA.  This means that without
bringing in additional information, such as convincing identifications
of very weak optical He~I lines (cf. Clocchiatti et~al. 1996, 1997) or
rigorous NLTE calculations of line optical depths for a hydrodynamical
model whose spectrum matches the entire observed spectrum (cf. Baron
et~al. 1996, 1999), there is some ambiguity concerning the
identification of the infrared absorption.  It may be primarily a
blend of He~I and C~I, but at present we cannot exclude the
possibility that it is partly, or even entirely, produced by Si~I.

Figure~5 compares the composite $+11$ day {\sl HST} spectrum and $+10$ day
optical spectrum to a synthetic spectrum that has $v_{phot}=9500$
\kms.  Now C~II is detached at 15,000 \kms.  In addition to the ions
needed for the optical spectrum, we have introduced Mg~II and Cr~II
for the UV.  More ions would need to be introduced to account for all
of the structure in the observed UV, but we leave detailed fitting of
the {\sl HST} UV for future work.

Figure~6 compares the $+26$ day observed spectrum to synthetic spectra
that have $v_{phot}=7000$ \kms.  In the upper panel the fit is good,
except (1) the synthetic O~I feature at 8900 \AA\ is too strong, (2)
the observed strong net emission in the Ca~II infrared triplet cannot
be produced with the resonance--scattering source function that we are
using, and (3) there is something missing from the synthetic spectrum
around 7000~\AA.  In the lower panel, lines of O~II are
introduced. The only ones that make a difference are the forbidden
lines [O~II] $\lambda\lambda$7320,7330, which we (Fisher et~al. 1999)
have discussed in connection with the peculiar Type~Ia SN~1991T.  In
SN~1991T our only alternative to introducing the [O~II] lines was to
lower our continuum placement and accept the observed flux minimum
near 7000~\AA\ as merely a gap between emissions.  We would not
necessarily expect the same gap to appear in the SN~Ia 1991T and the
SN~Ic 1994I, so the evidence for the presence of the [O~II] lines in
SN~1994I strengthens the case for their presence in SN~1991T.  We
regard the identification of the [O~II] lines in SN~1994I to be
probable, but not definite.

Figure~7, for the $+26$ day spectrum, is analogous to Figure~4 for the
$+7$ day spectrum.  In the top panel He~I, detached at 15,000 \kms,
accounts for the infrared absorption, but He~I $\lambda5876$ is too strong
in the synthetic spectrum.  In the middle panel undetached C~I and
He~I detached at 15,000 \kms\ combine to account for the IR
absorption.  In the lower panel, Si~I detached at 14,000 \kms\
accounts for the infrared absorption on its own.

\subsection{From $+4$ to $-4$ days}

Now we return to earlier epochs.  Figure~8 compares the $+4$ day
observed spectrum with a synthetic spectrum that has $v_{phot}=11,000$
\kms.  At this epoch we use a combination of Si~II, undetached but
with a maximum velocity of 12,000 \kms, and C~II detached at 15,000
\kms\, to account for the absorption near 6200~\AA.  Lines of Sc~II
and C~I also have been introduced, but the fit would be much the same
without them and we don't consider their identification to be
established.  Figures~9 and 10 compare the $-2$ and $-4$ day observed
spectra with synthetic spectra that have $v_{phot}=16,500$ and 17,500
\kms\ and maximum Si~II velocities of 18,000 \kms\ and 20,000 \kms,
respectively. Now lines of Mg~II also have been introduced but the
identification is not definite.  One or more additional ions would
need to be included to fit these spectra at wavelengths longer than
9000 \AA.

\subsection {The Inferred Composition Structure}

Figure~11 shows our constraints on the composition structure as
obtained from the optical spectrum.  The arrows show the velocity
intervals within which we think the ions have been detected in the
observed spectra.  The only ion on which we imposed a maximum velocity
was Si~II, around 20,000 \kms.  The only ion that we detached was
C~II, around 15,000 \kms; this may represent the minimum velocity of
the ejected carbon.

Because of the ambiguity of the identification of the infrared absorption,
He~I, C~I, and Si~I are not shown in Figure~11.  When He~I
$\lambda10830$ has been invoked to account for part of the IR
absorption, it has been detached at 15,000 or 18,000 \kms; thus if
helium is present at all, this is our estimate for its minimum
velocity.  For comparison, we note that in model CO21 of Iwamoto
et~al. (1994) the minimum carbon velocity is at about 12,000 \kms\ and
that of helium is at about 14,000 \kms.  In model 7A of Woosley,
Langer, \& Weaver (1995), which is favored by Woosley \& Eastman
(1997) for SN~1994I, the minimum velocities of helium and carbon are
lower, about 6000 and 8000 \kms, respectively.

A problem with attributing the infrared absorption to a blend of detached
He~I $\lambda$10830 and undetached C~I $\lambda$10695 is that using
undetached C~I down to 7000 \kms\ probably is inconsistent with using
C~II detached at 15,000 \kms\ for the optical spectrum, because carbon
ionization is unlikely to increase outward.  On the other hand, when
Si~I has been invoked to account for the infrared absorption it has been
detached at 14,000 \kms.  This is not necessarily inconsistent with
our use of Si~II between 10,000 and 20,000 \kms\ for the optical
spectrum if, as expected, silicon ionization decreases outward.

Figure~12 shows SYNOW plots in the infrared region, for four
ions that can produce strong absorption features near the observed IR
absorption.  This figure shows that observed infrared spectra of
SNe~Ic would help to resolve the true identity of the infrared absorption;
e.g., if Si~I is responsible for the infrared absorption then we would
expect another conspicuous Si~I absorption near 11,700 \AA.

\subsection {Mass and Kinetic Energy}

Our adopted values of $v_{phot}$ can be used to make rough estimates
of the mass and kinetic energy above the photosphere as a function of
time.  It is easily shown that for spherical symmetry and an $r^{-n}$
density distribution, the mass (in $M_\odot$) and the kinetic energy
(in $10^{51}$ ergs) above the electron--scattering optical depth
$\tau_{es}$ are

$$ M = (1.2 \times 10^{-4})\ v_4^2\ t_d^2\ \mu_e\ {n-1 \over n-3}\ 
\tau_{es}, $$

$$ E = (1.2 \times 10^{-4})\ v_4^4\ t_d^2\ \mu_e\ {n-1 \over n-5}\ 
\tau_{es}, $$

\noindent where $v_4$ is $v_{phot}$ in units of 10,000 \kms, $t_d$ is
the time since explosion in days, and $\mu_e$ is the mean molecular
weight per free electron.  For illustration we use $n=8$, the value we
used for the synthetic spectrum calculations, but it should be noted
that the value of $n$ is not well constrained by our fits.  We assume
that the explosion occurred on March 30, i.e., that the rise time to
maximum brightness was 9 days, and that $\mu_e = 14$ (e.g., a mixture
of singly ionized carbon and oxygen).  At $-4$ days, using $v_4=1.75$
and $\tau_{es}=2/3$ leads to $M = 0.1\ M_\odot$ and $E = 0.6 \times
10^{51}$ ergs.  At $+26$ days, $v_4=0.7$ and $\tau_{es}=2/3$ give
$M=0.9\ M_\odot$ and $E=0.8 \times 10^{51}$ ergs.  The latter values
may be overestimates of the values above the photosphere at $+26$ days
because at this epoch the photosphere could be at $\tau_{es} < 2/3$
owing to the contribution of lines to the pseudo--continuous opacity.
Of course, there is additional mass (but not much kinetic energy)
beneath the photosphere.  In any case, our spectroscopic estimates of
the mass and kinetic energy of SN~1994I are not inconsistent with
previous estimates based on light--curve studies.

\section{Discussion}

The photospheric--phase optical spectra of SN~1994I can be fit rather
well by SYNOW synthetic spectra that include only ``reasonable'' ions,
i.e., those that can regarded as candidate ions based on LTE
optical--depth calculations (Hatano et~al. 1999).  In addition to the
obvious Ca~II, Na~I, O~I, and Fe~II, we have invoked Ti~II, C~II,
Si~II, and [O~II], and at the earliest phases, C~I, Sc~II, and Mg~II.
We regard the identification of Ti~II to be definite.  Although it can
be difficult to distinguish Si~II from detached C~II, we think that
both Si~II and C~II are needed.  We regard the presence of [O~II],
Sc~II, and Mg~II to be probable.

It is difficult to account for the infrared absorption with He~I
$\lambda10830$ alone.  It may be a blend of C~I $\lambda10695$ and
He~I $\lambda$10830, but then there is an apparent inconsistency
between the minimum velocities of carbon as inferred from C~I and
C~II.  Alternatively, the entire infrared feature can be fit by lines of
Si~I without compromising the fit in the optical. The true
identification of the infrared absorption is a matter that can only be
decided by means of detailed NLTE spectrum calculations for realistic
hydrodynamical models whose emergent spectra closely match the rest of
the optical and infrared spectrum.

The spectra of SN~1994I can be well fit with a simple model that
includes the assumption of spherical symmetry.  Thus, we certainly are
not forced to invoke an asymmetry.  However, the flux spectrum, unlike
the polarization spectrum, is not very sensitive to a mild asymmetry
(Jeffery \& Branch 1990; H\"oflich et~al. 1996), so the degree to
which we can assert that SN~1994I was {\sl not} very asymmetric
(cf. Wang \& Wheeler 1998) is unclear.  We intend to investigate this
issue in future work.

Our spectroscopic estimates of the kinetic energy carried by matter
above the photosphere are consistent with previous estimates based on
the light curve, i.e., they are near the canonical supernova value of
$10^{51}$ ergs.  In this respect, SN~1994I is valuable for comparison
with the peculiar SNe~Ic 1998bw and 1997ef.  We find that the same
spectroscopic procedure for estimating the kinetic energy gives much
higher kinetic energies for SNe ~1997ef and 1998bw (Branch 1999;
Millard et~al. in preparation).  This leads us to believe that
SNe~1997ef and 1998bw were hyper--energetic.

This work has been supported by NSF grants AST-9417102 to D.B.,
AST-9731450 to E.B., AST-9417213 to A.V.F., and NASA GO--2563.001 to
the SINS group from the Space Telescope Science Institute, which is
operated by AURA under NASA contract NAS 5--26555.

\clearpage

\begin{figure}
\figcaption{Spectra of SN~1994I that have been compared to SYNOW
synthetic spectra.  Epochs are in days with respect to the time of
maximum brightness in the $B$ band, 8~April 1994~UT.  Panel~(a):
spectra obtained at the Lick Observatory, from Filippenko
et~al. (1995), with an {\sl HST} spectrum obtained by the SINS group
at $+11$ days attached to the Lick spectrum obtained at $+10$ days. (A
5 \AA\ boxcar smoothing has been applied to the {\sl HST} spectrum.)
Panel~(b): spectra obtained by Brian Schmidt and R.P.K. at the MMT.
Telluric absorption near 6860 and 7600 \AA\ has been removed from the
Lick spectra, but not from the MMT spectra.  In this and subsequent
figures the flux is per unit frequency and wavelengths in the
supernova rest frame, for $cz=461$ \kms. The scale of the vertical
axis is arbitrary.
\label{fig1}}
\end{figure}

\begin{figure}
\figcaption{ The velocity at the photosphere, $v_{phot}$, adopted in
our SYNOW synthetic spectra, is plotted against time.
\label{fig2}}
\end{figure}

\begin{figure}
\figcaption{ A spectrum of SN~1994I obtained at $+7$ days is compared
to a synthetic spectrum that has $v_{phot}=10,000$ \kms.  C~II is
detached at 16,000 \kms.  Ions responsible for features in the synthetic
spectra are indicated near the absorption components of the calculated
features.
\label{fig3}} 
\end{figure}

\begin{figure}
\figcaption{A spectrum of SN~1994I obtained at $+7$ days is compared
to three synthetic spectra that have $v_{phot}=10,000$ \kms.  To
account for the observed infrared feature, the synthetic spectra include
lines of He~I detached at 15,000 \kms\ (top panel); undetached C~I and
He~I detached at 18,000 \kms\ (middle panel); and Si~I detached at
14,000 \kms\ (bottom panel).
\label{fig4}}
\end{figure}

\begin{figure}
\figcaption{A composite spectrum of SN~1994I consisting of the $+11$
day {\sl HST} spectrum and a $+10$ day optical spectrum is compared to a
synthetic spectrum that has $v_{phot}=9500$ \kms.  C~II is detached at
15,000 \kms.
\label{fig5}}
\end{figure}

\begin{figure}
\figcaption{ A spectrum of SN~1994I obtained at $+26$ days is compared
to two synthetic spectra that have $v_{phot}=7000$ \kms.  The
synthetic spectrum in the lower panel is like that of the the upper
panel except that lines of [O~II] are introduced.
\label{fig6}} 
\end{figure}

\begin{figure}
\figcaption{A spectrum of SN~1994I obtained at $+26$ days is compared
to three synthetic spectra that have $v_{phot}=7000$ \kms.  To account
for the observed infrared feature, the synthetic spectra include lines of
He~I detached to 15,000 \kms\ (top panel); undetached C~I and He~I
detached to 15,000 \kms\ (middle panel); and Si~I detached to 14,000
\kms\ (bottom panel).
\label{fig7}}
\end{figure}

\begin{figure}
\figcaption{ A spectrum of SN~1994I obtained at $+4$ days is compared
to a synthetic spectrum that has $v_{phot}=11,000$ \kms.
\label{fig8}} 
\end{figure}

\begin{figure}
\figcaption{ A spectrum of SN~1994I obtained at $-2$ days is compared
to a synthetic spectrum that has $v_{phot}=16,500$ \kms.
\label{fig9}} 
\end{figure}

\begin{figure}
\figcaption{ A spectrum of SN~1994I obtained at $-4$ days is compared
to a synthetic spectrum that has $v_{phot}=17,500$ \kms.
\label{fig10}} 
\end{figure}

\begin{figure}
\figcaption{ A summary of the velocity intervals within which we
think the ions affect the observed spectra; see text.
\label{fig11}}
\end{figure}

\begin{figure}
\figcaption{SYNOW spectra ($v_{phot}=10,000$ \kms) of four ions that
can produce strong absorption features near the observed infrared feature
\label{fig12}}
\end{figure}

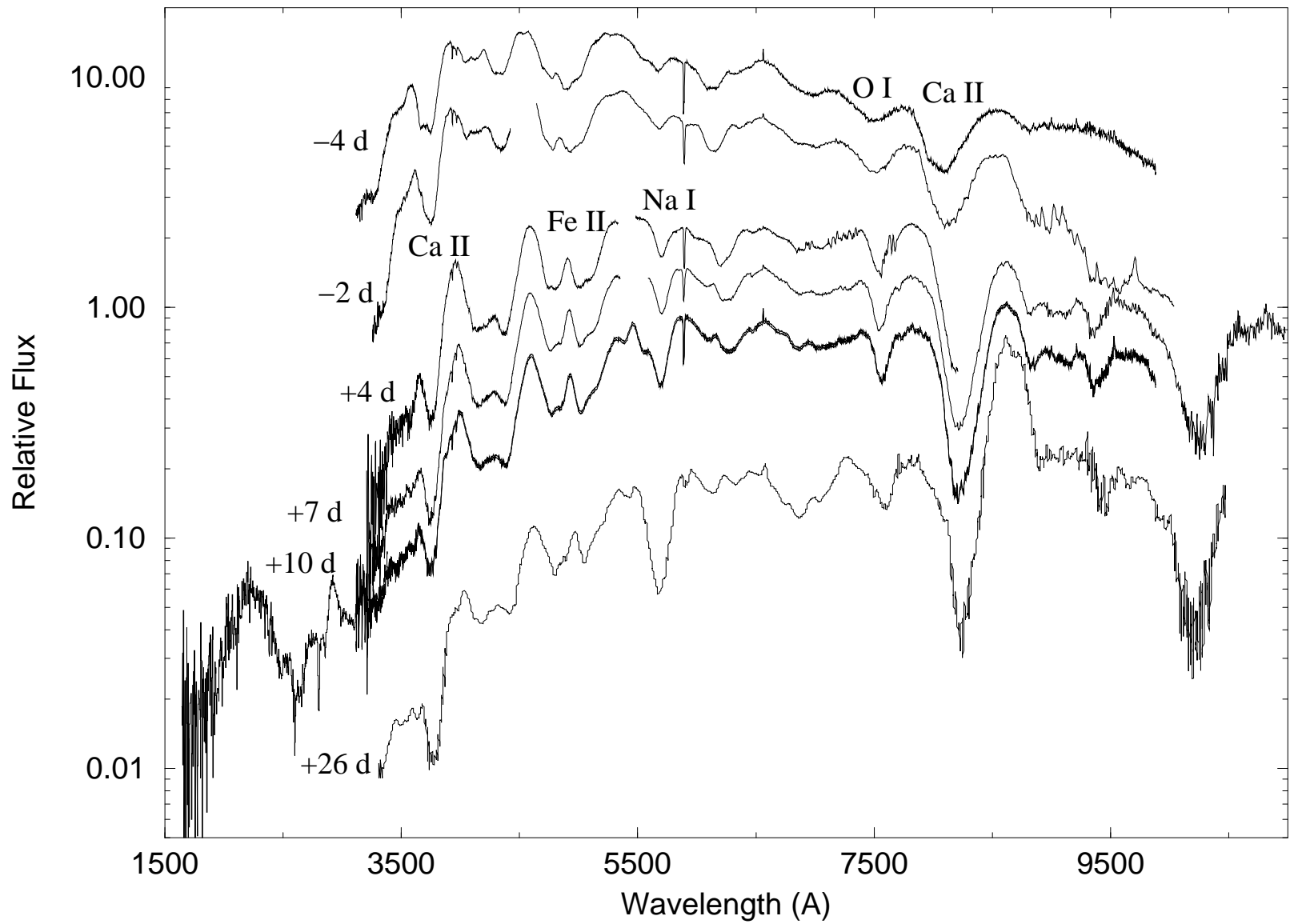
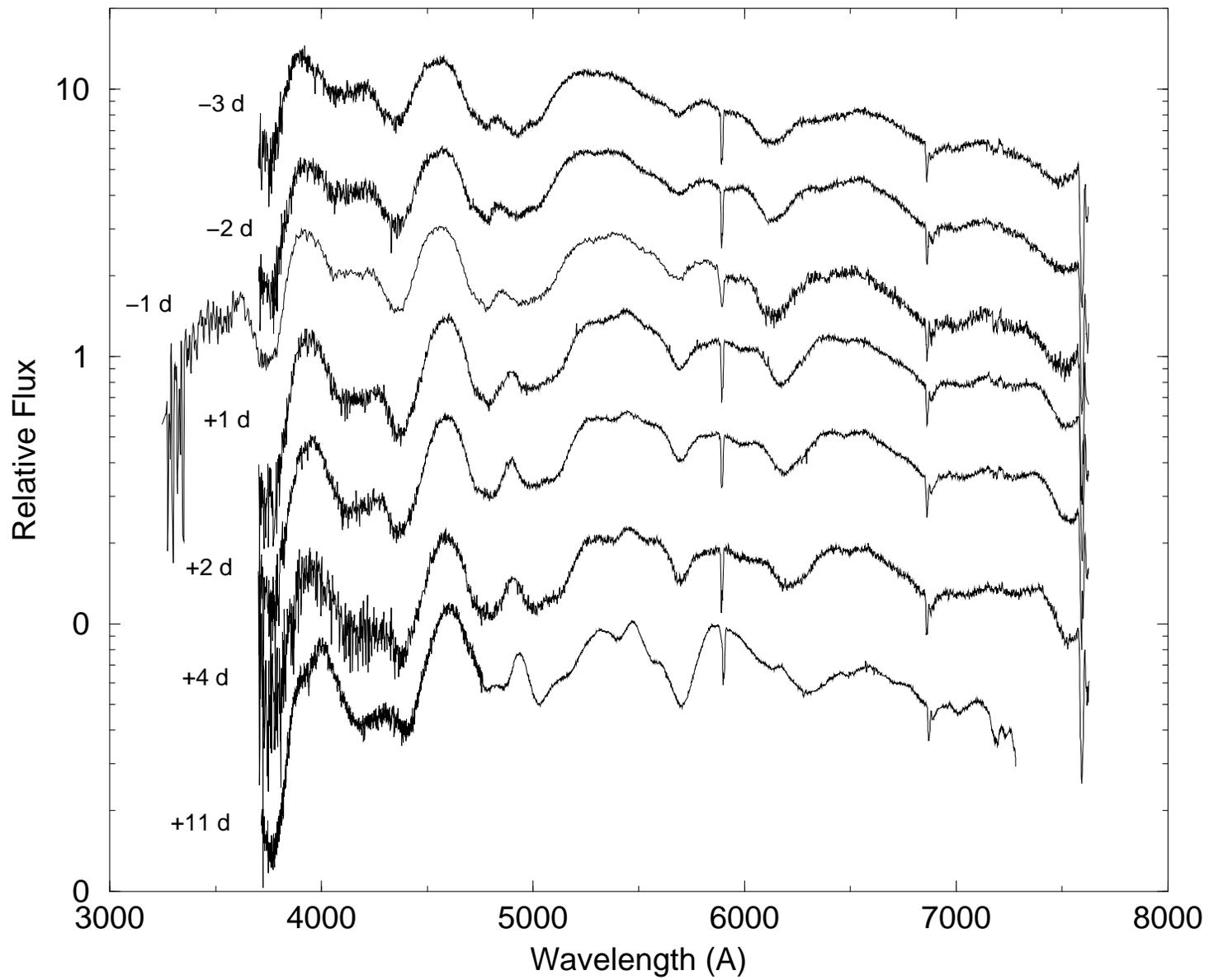
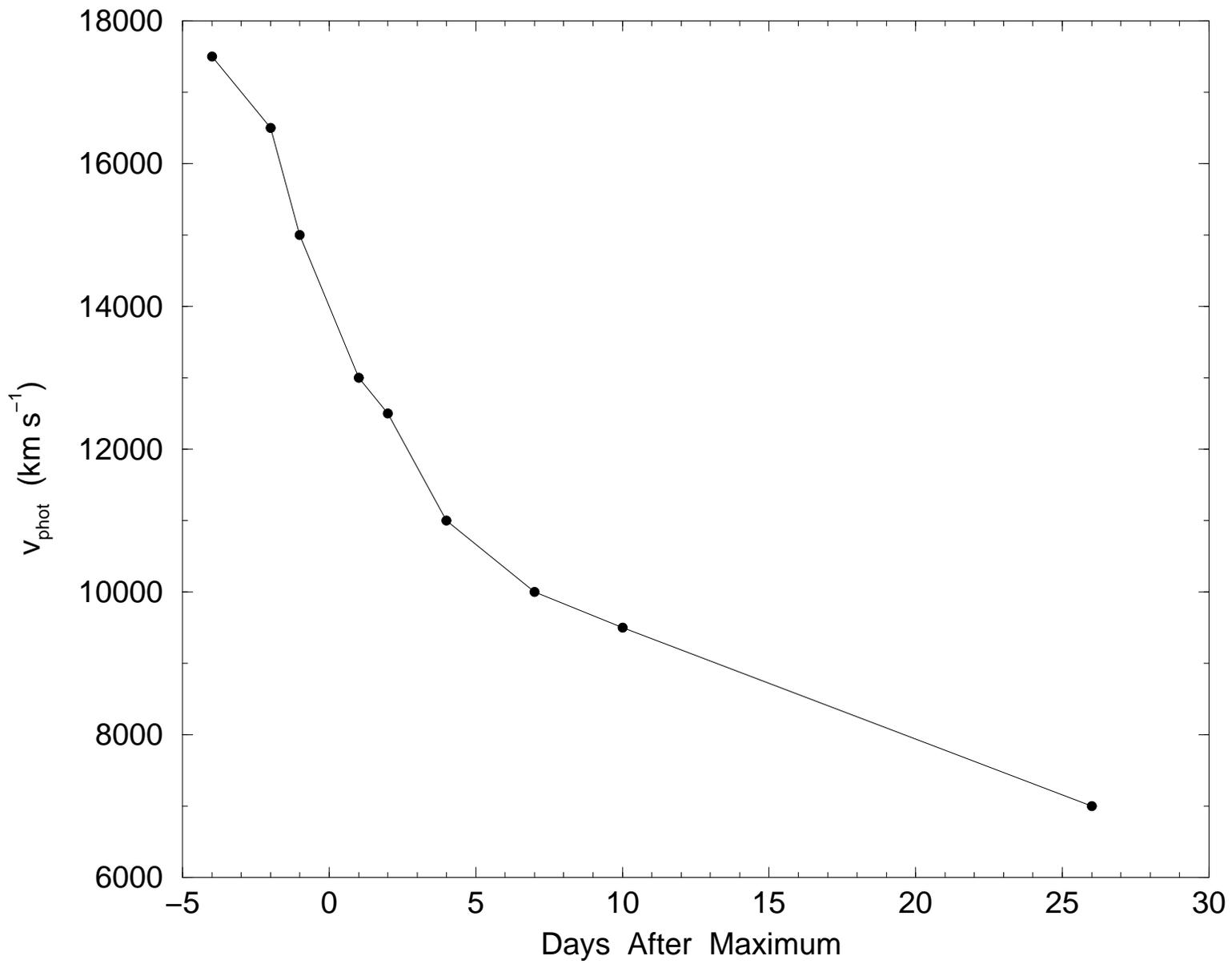
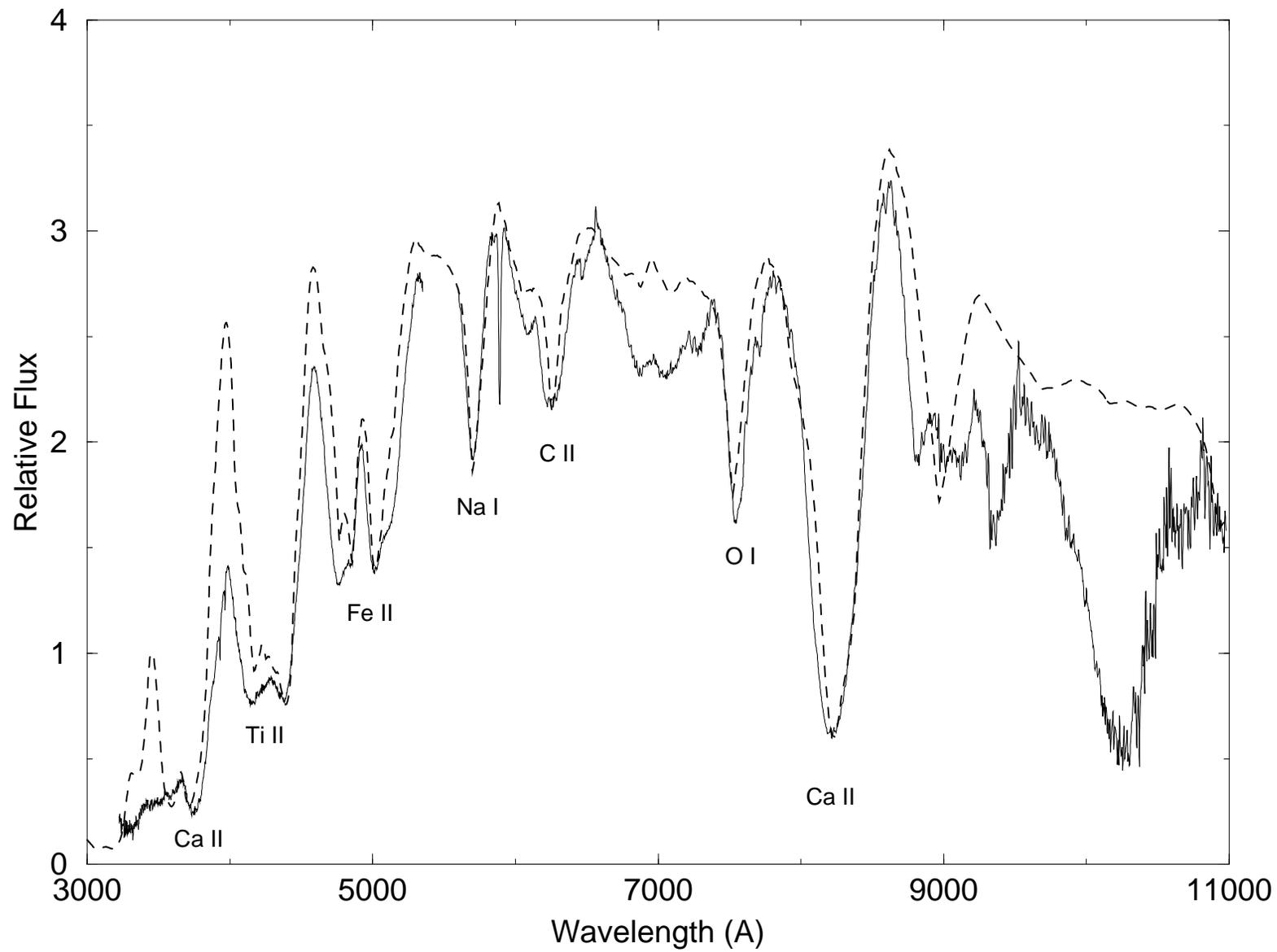
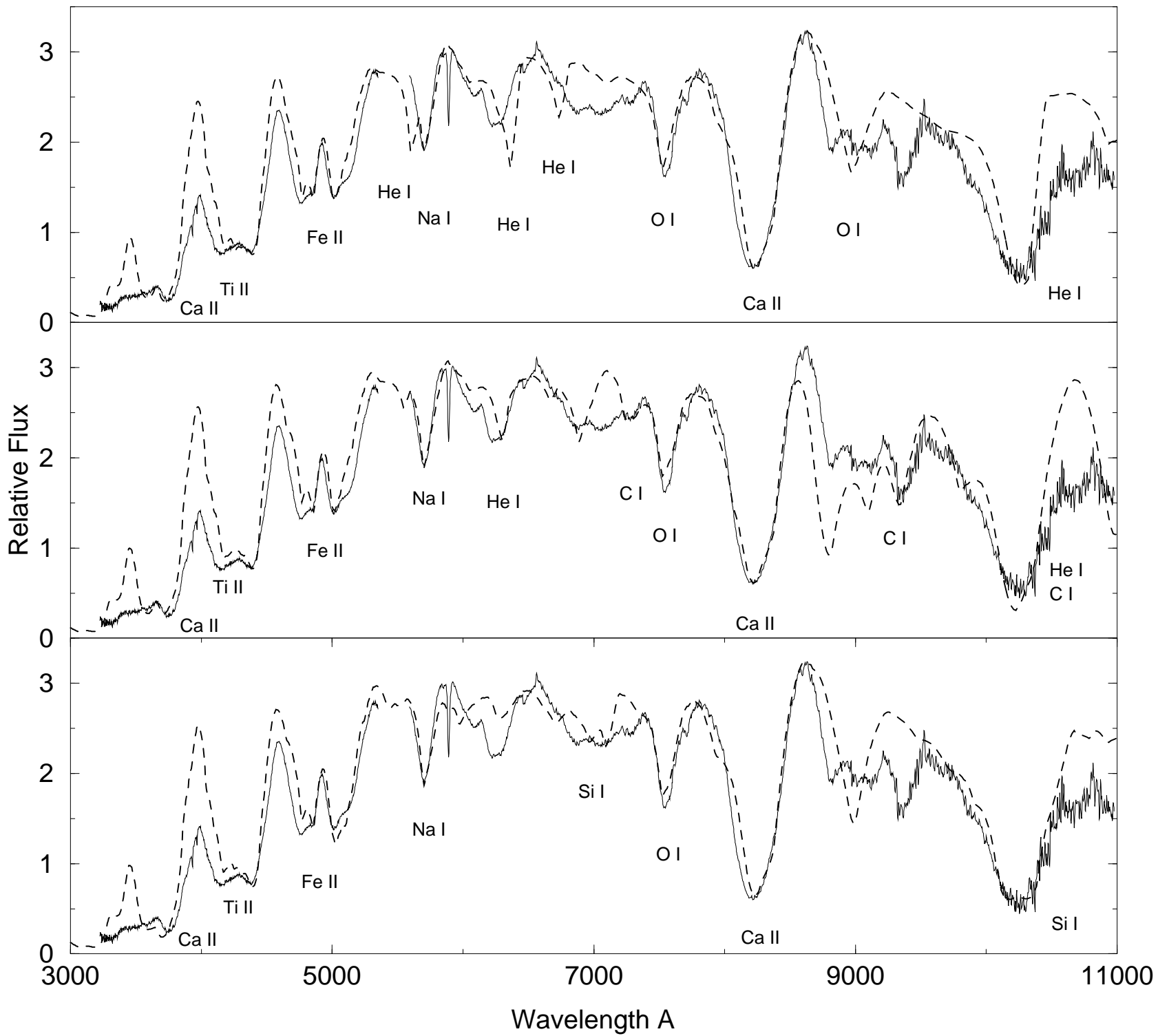
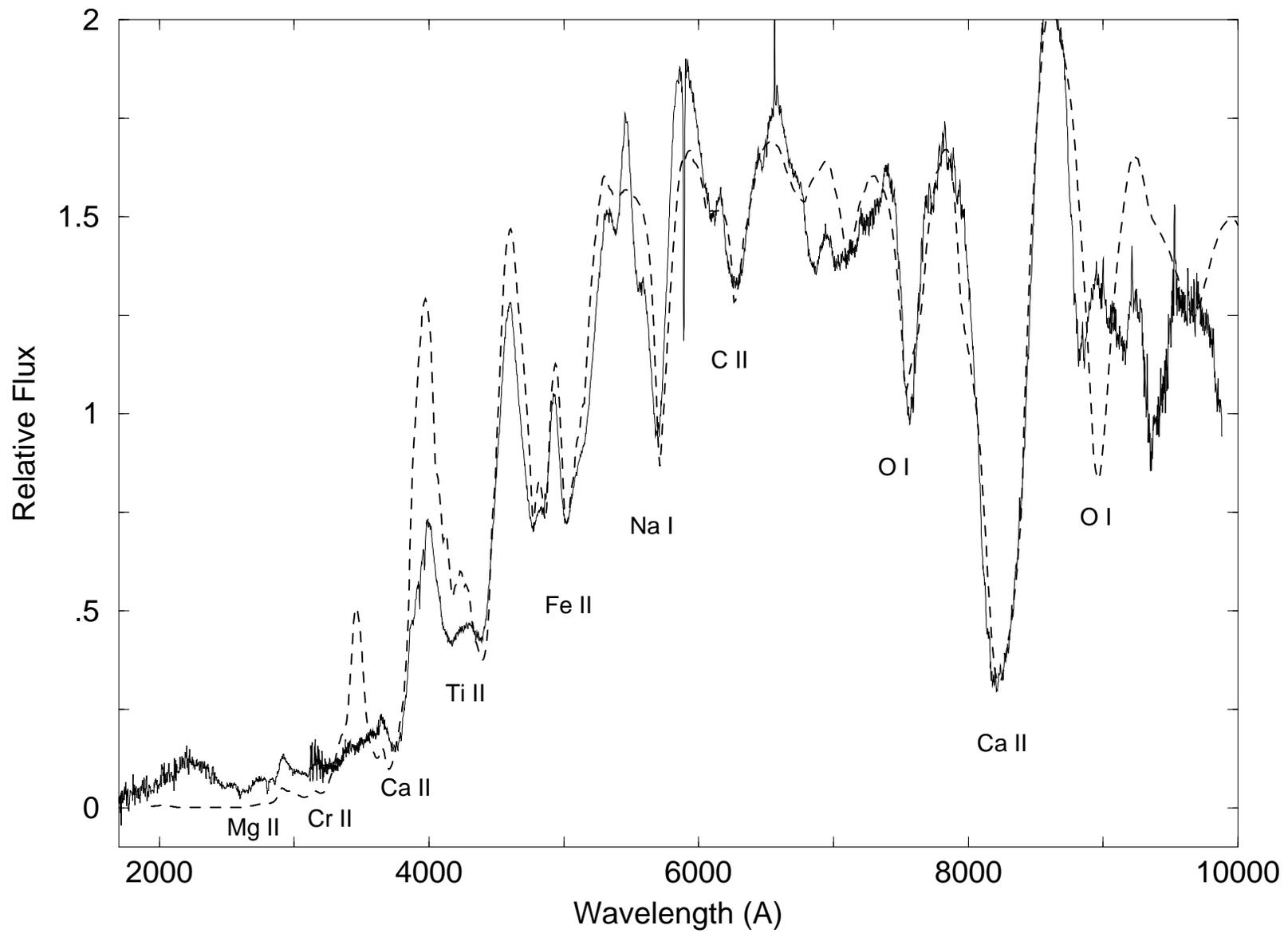
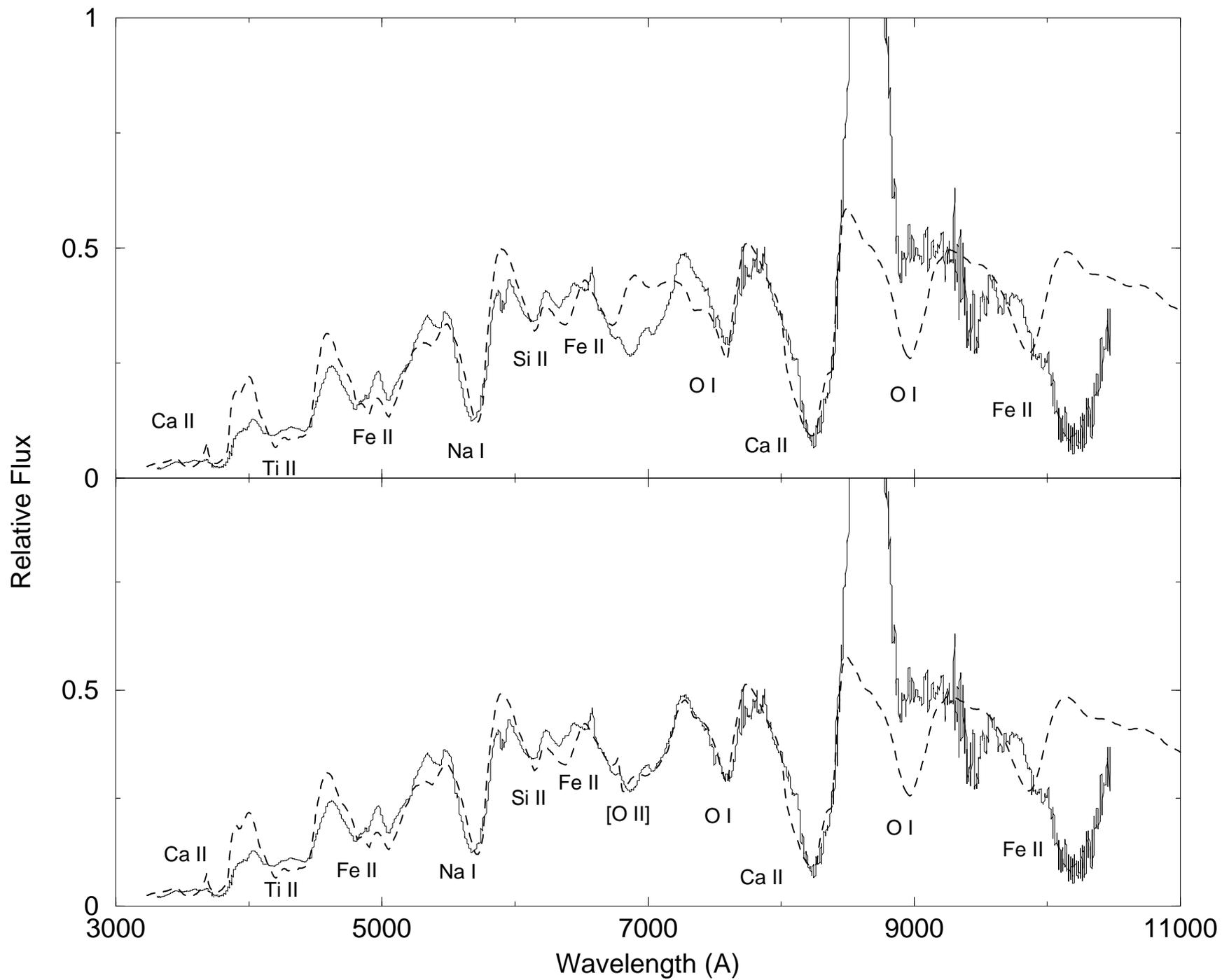
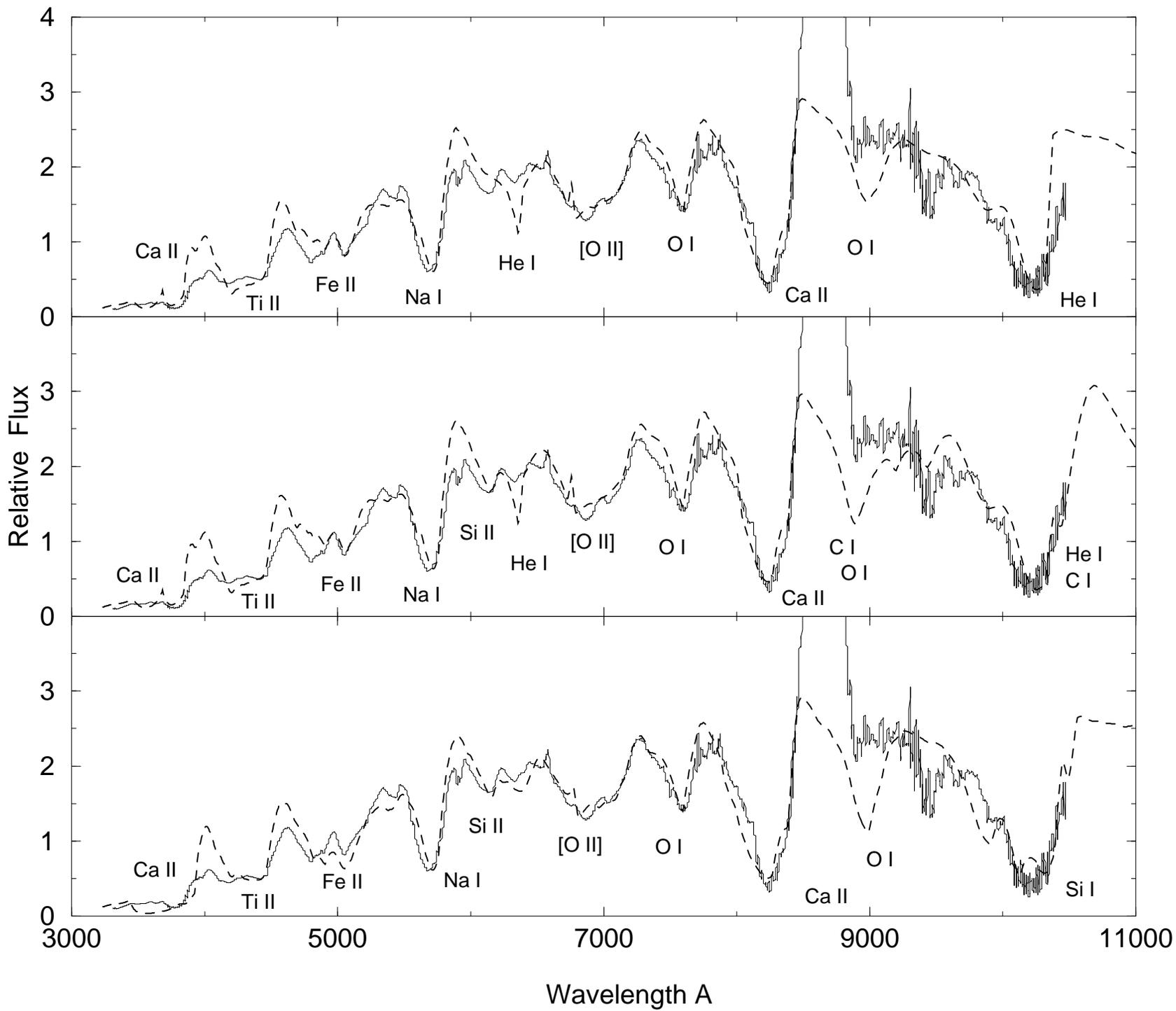
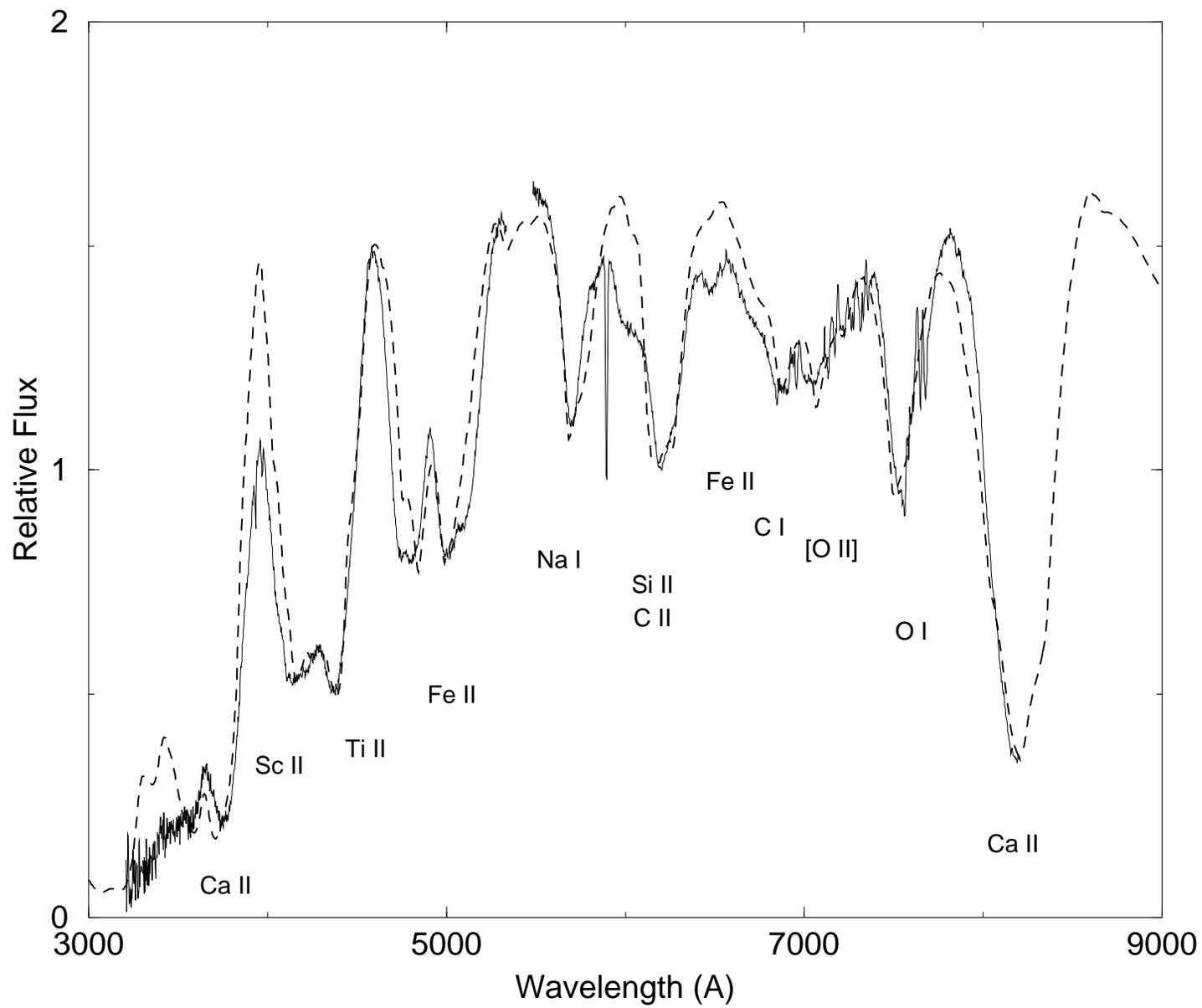
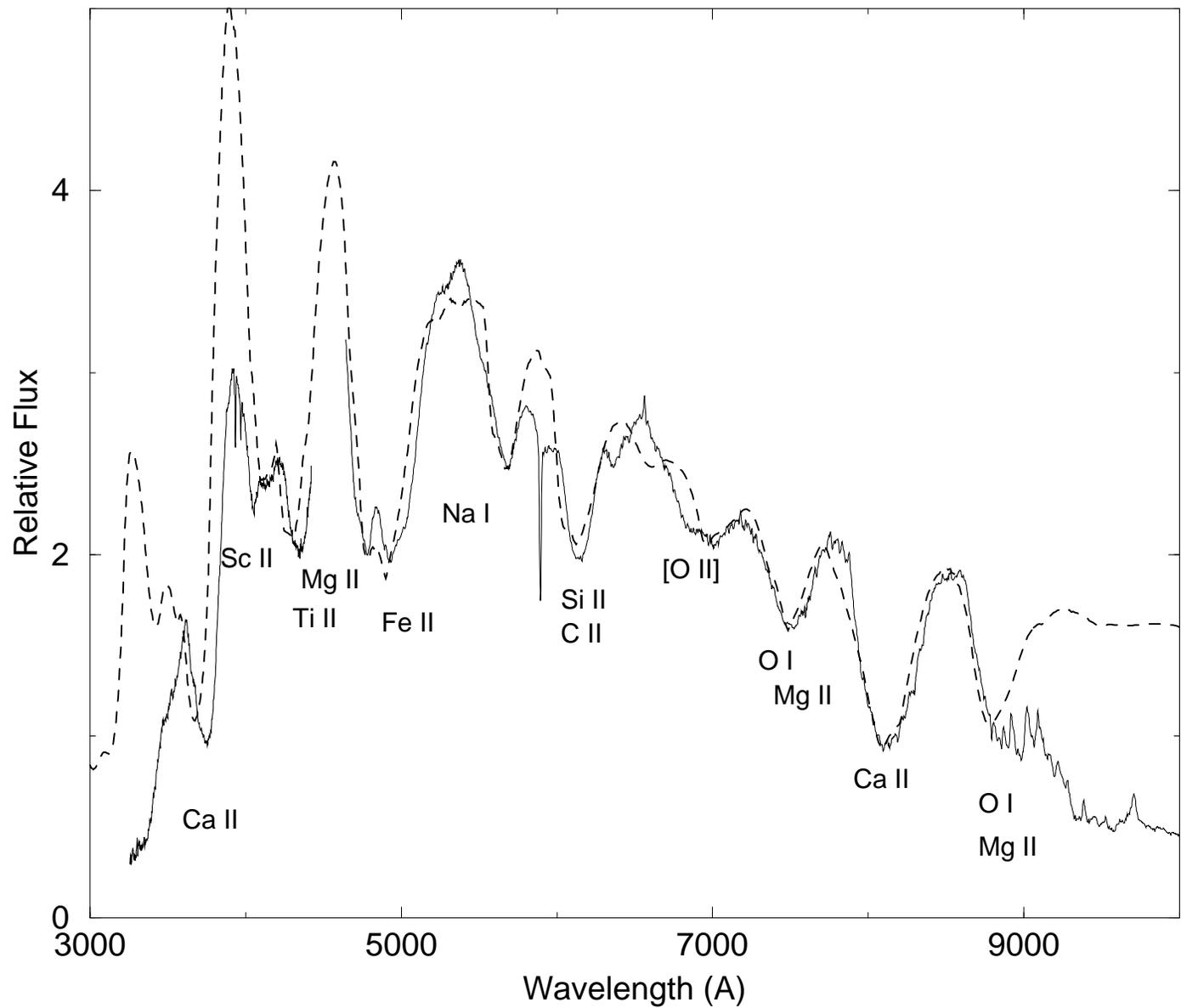
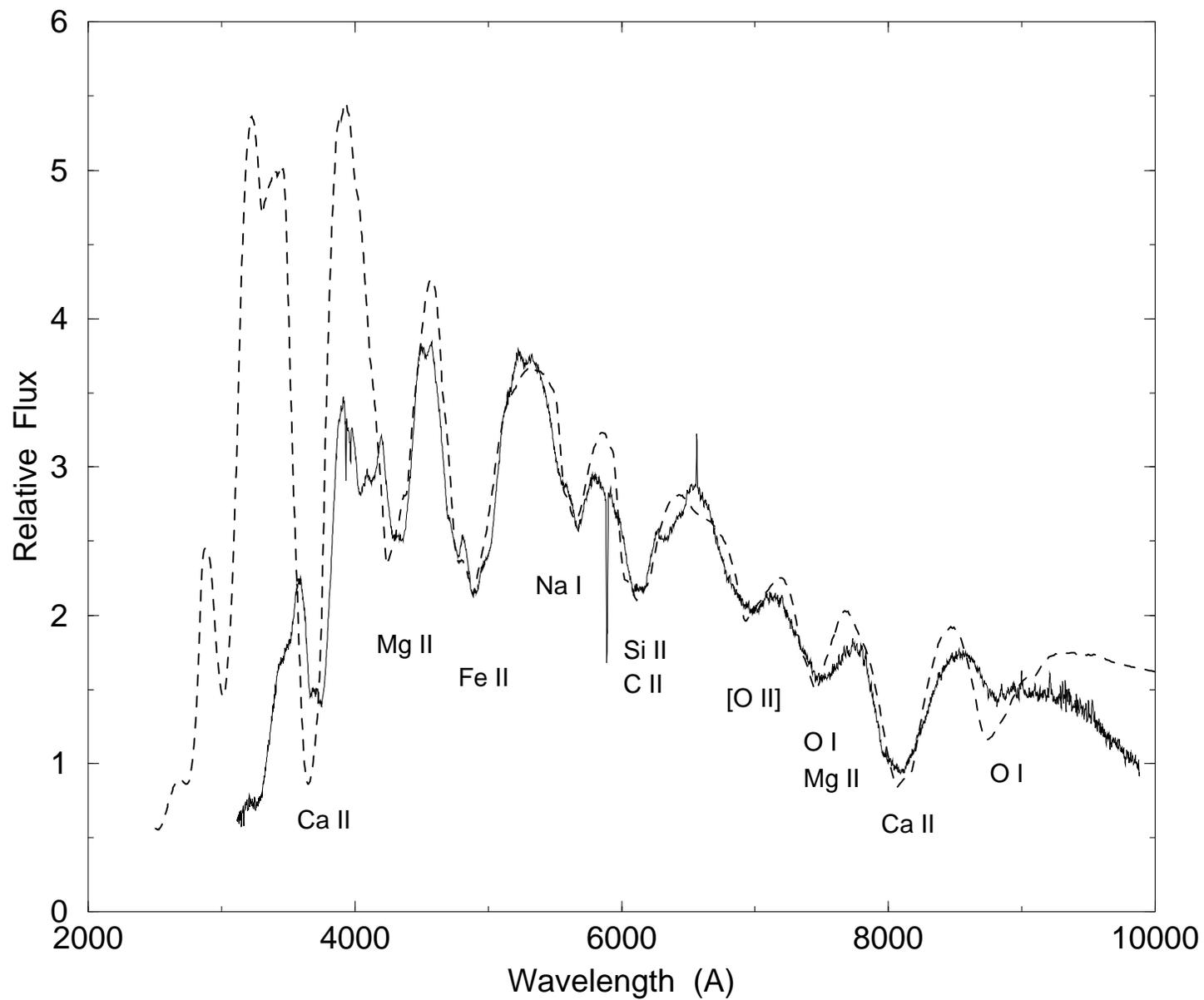
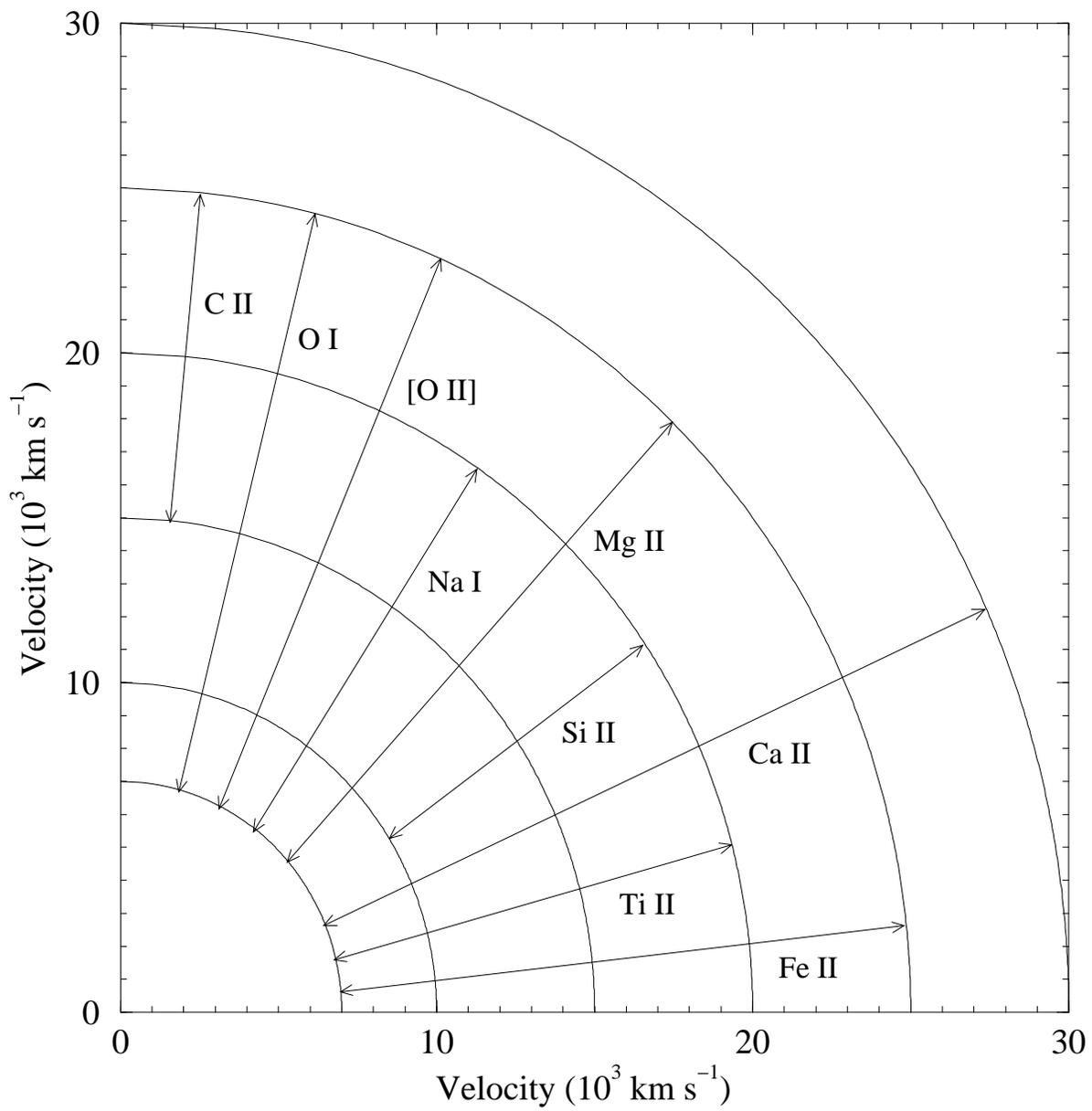
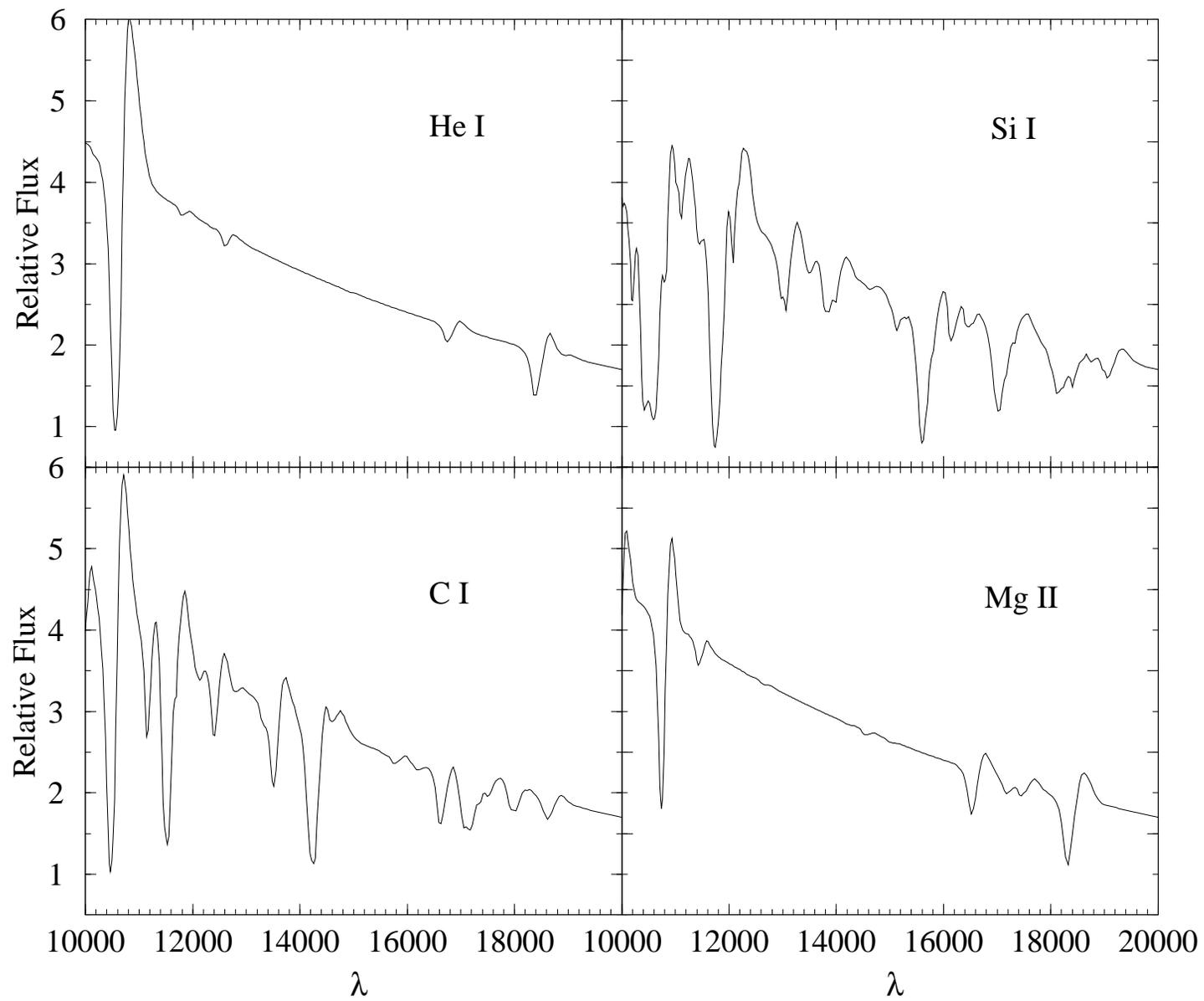
\end{document}